\documentclass[aps,prl,twocolumn,amsmath,amssymb]{revtex4}
\usepackage{graphicx}% Include figure files
\usepackage{dcolumn}% Align table columns on decimal point
\usepackage{bm}% bold math

\begin{document}

\draft
\title{Comment on ``Drip Paintings and Fractal Analysis'' by K. Jones-Smith, H. Mathur and L.M. Krauss (arXiv:0710.4917v2 [cond-mat.stat-mech])}
\author{A.P. Micolich$^{1}$}
\email{mico@phys.unsw.edu.au}
\author{B.C. Scannell$^{2}$}
\author{M.S. Fairbanks$^{2}$}
\author{T.P. Martin$^{1}$}
\author{R.P. Taylor$^{2}$}
\email{rpt@uoregon.edu}

\affiliation{1. School of Physics, University of New South Wales,
Sydney NSW 2052, Australia}

\affiliation{2. Physics Department, University of Oregon, Eugene OR
97403-1274, U.S.A.}

\date{\today}

\maketitle

In a recent manuscript, Jones-Smith \emph{et al.} attempt to use the
well-established box-counting technique \cite{Gouyet96} for fractal
analysis to ``\emph{demonstrate conclusively that fractal criteria
are not useful for authentication}''\cite{Jones-Smith07}. Here, in
response to what we view to be an extremely simplistic
misrepresentation of our earlier work \cite{TaylorPRL07} by
Jones-Smith \emph{et al.}, we reiterate our position regarding the
potential of fractal analysis for artwork authentication. We also
point out some of the flaws in the analysis presented in
\cite{Jones-Smith07} by Jones-Smith \emph{et al.}

We begin by reiterating our position regarding the potential of our
analysis \cite{TaylorPRL07} for the authentication of artwork.
Jones-Smith \emph{et al.} state that ``\emph{It has been claimed
[1-6] that fractal analysis can be applied to unambiguously
characterize works of art such as the drip paintings of Jackson
Pollock}'', ``\emph{...that all Pollock drip paintings satisfy these
criteria [in their Ref. 6]}'', and that ``\emph{...these
characteristics [in their Ref. 6] are exclusive to
Pollock}''\cite{Jones-Smith07}. These are misrepresentations of our
recent paper \cite{TaylorPRL07} that Jones-Smith \emph{et al.} use
to portray our work as a `black-box' authenticator, where an image
is simply fed in one end and a final verdict of authentic or
non-authentic pops out the other. This would be an amateur and
superficial approach to authentication, akin to trying to solve a
complex detective case with only one clue.

We strongly discourage this simplistic `black-box' view of our
results. In particular, we have never stated that the presence of
these signatures proves that a painting is by Pollock, nor that
their absence indicates a non-Pollock. We've merely pointed out that
the presence or absence of fractal content in an artwork may be
useful as one of the many clues needed to make an informed decision
regarding the authenticity of an artwork. We have published one
article regarding authenticity issues and it is very clear on this
point \cite{TaylorPRL07}. This is further reinforced in Abbott's
article \cite{Abbott07} where Taylor is quoted directly as saying
``\emph{Taken in isolation, these results are not intended to be a
technique for attributing a poured painting to Jackson Pollock.
However, the results may be useful when coupled with other important
information such as provenance, connoisseurship and materials
analysis.}''

Looking now at two prominent flaws in this latest work by
Jones-Smith \emph{et al.} [1], we firstly note that their
box-counting analysis regularly produces fractal dimensions $D > 2$
(e.g., for the composite layer in Untitled 1, white layer in
Untitled 14, all layers in Free-form, black layer in Number 8, both
layers in Composition with red and black). This is an impossible
result in a correctly written box-counting algorithm
\cite{Gouyet96}, which should always produce $1 \leqslant D
\leqslant 2$ when applied to two-dimensional images such as the
artwork considered by Jones-Smith \emph{et al.}
\cite{Jones-Smith07}. This points to a serious error in their
box-counting analysis that causes it to produce incorrect counting
statistics, and thereby, mathematically impossible results.

Secondly, we note that one of the key objections of Jones-Smith
\emph{et al.} to our work is that ``\emph{it is mathematically
impossible for the visible portion of each layer and the composite
to separately behave as fractals in a multilayered painting.}''
\cite{Jones-Smith07}, a claim that was also made in their earlier
paper \cite{Jones-Smith06} and rebutted in our response
\cite{TaylorNat06}. Given this, it is interesting to note the
box-counting results obtained by Jones-Smith \emph{et al.} for
Pollock's `The Wooden Horse' (Number 10A, 1948)
\cite{Jones-Smith07}. In particular, they analyze the blue and black
layers separately, and together as a composite (the PS result) and
they find that all three are fractal. According to their earlier
claims \cite{Jones-Smith06}, this outcome is a mathematical
impossibility, which either proves that their box-counting algorithm
is producing incorrect results, or that their argument regarding the
fractality of composites \cite{Jones-Smith06} is false. We
conjecture that both are occurring, because although the results
above show their box-counting algorithm produces incorrect results,
our box-counting algorithm also shows that two layers and their
composite can all show fractal behavior in certain cases. A more
thorough presentation of the flaws in the work of Jones-Smith
\emph{et al.} will be the subject of a future publication.

In summary, we have highlighted two significant mathematical and
logical flaws in the recent work by Jones-Smith \emph{et al.}
\cite{Jones-Smith07}, such that the results should be treated with
skepticism. We also reiterate that our recent work
\cite{TaylorPRL07} is meant to contribute useful information towards
the wealth of knowledge required to authenticate an artwork, it is
not meant to be a `black-box' authenticator as Jones-Smith \emph{et
al.} attempt to portray in their work
\cite{Jones-Smith07,Jones-Smith06}.


\begin{thebibliography}:

\bibitem{Jones-Smith07} K. Jones-Smith, H. Mathur and L.M. Krauss, arxiv:0710.4917v2 [cond-mat.stat-mech] (2007).

\bibitem{Gouyet96} J.-F.Gouyet, ``Physics and Fractal Structures'', (Springer-Verlag, New York, 1996).

\bibitem{TaylorPRL07} R.P. Taylor, R. Guzman, T.P. Martin, G.D.R. Hall, A.P. Micolich, D. Jonas, B.C. Scannell, M.S. Fairbanks and C.A. Marlow, Patt. Recog. Lett. {\bf28}, 695 (2007).

\bibitem{Abbott07} A.Abbott, Nature {\bf439}, 648 (2006).

\bibitem{Jones-Smith06} K. Jones-Smith and H. Mathur, Nature {\bf444}, E9 (2006).

\bibitem{TaylorNat06} R.P. Taylor, A.P. Micolich and D. Jonas, Nature {\bf444}, E10 (2006).

\end{thebibliography}
\end{document}